\documentclass[aps,prb,twocolumn,amsmath,amssymb,nofootinbib,tighten,showpacs,floatfix]{revtex4}

\usepackage{pslatex}
\usepackage{graphicx}
\usepackage{dcolumn} % Align table columns on decimal point
\usepackage{bm}      % bold math

\begin{document}
\title{Universality of delocalization in unconventional
       dirty superconducting
       wires with broken spin-rotation symmetry}
\author{P.\ W.\ Brouwer}
\affiliation{Laboratory of Atomic and Solid State Physics,
             Cornell University, 
             Ithaca, 
             New York 14853-2501}
\author{A.\ Furusaki}
\affiliation{Yukawa Institute for Theoretical Physics,
             Kyoto University,
             Kyoto 606-8502, 
             Japan}
\author{C.\ Mudry}
\affiliation{Paul Scherrer Institute,
             CH-5232 Villigen PSI, 
             Switzerland \\
             and Yukawa Institute for Theoretical Physics,
             Kyoto University,
             Kyoto 606-8502, 
             Japan}
\date{\today}
\begin{abstract}
In dirty superconducting wires, quasiparticle states at the
Fermi level need not be
exponentially localized if spin-rotation symmetry is broken
[Brouwer {\em et al.}, Phys.\ Rev.\ Lett.\ {\bf 85}, 1064 (2000)].
Here we present evidence that not-localized states are generic in the
thick-wire limit, while for wires of finite thickness delocalization 
requires fine tuning of the disorder,
consistent with earlier results of Motrunich 
{\em et al.} [Phys.\ Rev.\ B {\bf 63}, 224204 (2001)].
The thick-wire limit is defined as the simultaneous
limit where the length $L$ of the wire and the number $N$ of
propagating channels at the Fermi energy are both taken to infinity
with their ratio held fixed.
\end{abstract}


\maketitle 

Localization properties of weakly disordered normal-metal wires 
depend first and foremost on the fundamental symmetries of
the wire: presence or absence of time-reversal symmetry and
spin-rotation symmetry. In addition to these, two more 
symmetries can play an important role in the characterization of
quantum wires: chiral symmetry and particle-hole symmetry. The
former is relevant for lattice models with randomness in the hopping
amplitudes only, while the latter applies to superconducting wires
or normal-metal wires in the proximity of a superconductor.

Is symmetry the only player determining the localization 
properties? For normal-metal wires, it is generally agreed on that
this is the case when the disorder is weak. For superconducting 
wires, different views have
been published in the literature. Advocates of field-theoretic, 
random-matrix, 
or Fokker-Planck approaches
have mostly considered the symmetry classification as sufficient,
thus arriving at a division of superconducting wires into four
universality 
classes.\cite{AltlandZirnbauer,HigherD,Bundschuh,Senthil,classD,Bocquet,BFGM}
These are labeled C, CI, D, and DIII, and are
characterized by the presence or absence of time-reversal symmetry
and spin-rotation symmetry.\cite{AltlandZirnbauer}
For the classes D and DIII, where spin-rotation symmetry is broken, 
it was found that wave functions are ``critical'' if disorder is weak, 
rather than exponentially localized, on length scales beyond a characteristic
crossover scale $N l$, where $N$ is the number of 
propagating channels in the wire and $l$ is the mean free
path. Here ``critical''  indicates algebraic
decay of the average conductance, opposed to exponential decay if
wave functions are localized.
Disorder is ``weak'' if $k_F l \gg 1$, $k_F$ being the Fermi wave vector.
On the other hand, using a numerical
transfer matrix method to calculate localization lengths in 
single-channel wires and strong-disorder renormalization-group
arguments, Motrunich {\em et al}.\ found that, if spin-rotation
symmetry is broken, the localization length 
only diverges if the disorder is fine tuned.\cite{Motrunich}
They concluded that
symmetry alone is not sufficient to characterize disordered wires
with particle-hole symmetry.
Such a finding is consistent with the expectation that, 
in a single-channel wire, weak spin-orbit scattering or weak superconducting
correlations cannot destroy the localized phase that exists 
in the presence of spin-rotation symmetry and without superconductivity.

In this paper we take a more detailed look at this apparent 
contradiction for the case of class D where, in addition to
spin-rotation symmetry, time-reversal symmetry is broken as well. 
We consider both regimes $L \gg N l$ and
$L \lesssim N l$. For thick wires ($N \gg 1$), the regime
$l \ll L \ll  N l$ corresponds to the diffusive regime.
For thin wires ($N$ of order unity), one has 
$N l\sim l$, and there is no diffusive regime. 
Our main conclusion --- to be supported by numerical
and analytical arguments below --- is that, as argued in
Ref.\ \onlinecite{Motrunich}, for quantum wires
of finite thickness and broken spin-rotation symmetry, 
fine tuning of the disorder is required in order
to obtain a diverging localization length. However, we shall 
also argue that in the limit $N \to \infty$  
of ``thick'' quantum wires symmetry alone is
sufficient to determine the localization properties.\cite{foot1} 
We note that this criterion is not different from that in 
normal wires.\cite{foot2}
Thus, as far as the classification based on symmetry is concerned, 
superconducting disordered wires are no less universal than ordinary
normal-metal wires.
In the final analysis, realization of the quasi-one-dimensional
critical behavior of class D demands fine tuning
of microscopic parameters for a wire of finite thickness but becomes
generic in the thick-wire limit provided the disorder is
sufficiently weak. We reach this conclusion by studying the competition
between localization in the standard unitary class (no superconductivity)
and criticality in class D that occurs upon increasing the
strength of superconducting correlations in the wire.
However, the same scenario applies to the crossover between the
localized phase of class C and the critical phase of class D 
upon breaking of the spin-rotation symmetry.

As in Ref.\ \onlinecite{BFGM}, our starting point is the Hamiltonian
\begin{equation}
{\cal H} = {\cal K} + {\cal V}, \qquad 
{\cal K} = 
  \sigma_0 \otimes \gamma_0 \otimes \tau_3 \otimes \openone_{N}
  \,i v_F \partial_x,
  \label{eq:cal H}
\end{equation}
where $\sigma_0$ is the $2 \times 2$ unit matrix in the spin grading,
$\gamma_0$ is the $2 \times 2$ unit matrix in particle-hole grading,
and $\tau_3$ is the Pauli matrix in left/right mover grading.
The kinetic energy ${\cal K}$ describes the propagation of right
and left moving quasiparticles in $N$ channels at the Fermi level. The
``potential'' ${\cal V}(x)$ is an $8N \times 8N$ matrix that accounts
for the presence of both disorder and superconducting
correlations. In particle/hole ($\gamma$) grading it reads
\begin{equation}
  {\cal V} =
\left(
\begin{array}{cc}
v
&
\Delta\\
-\Delta^*
&-v^{\rm T}
\end{array}
\right),
\label{eq:def V}
\end{equation}
where $v$ ($\Delta$) is a Hermitian (antisymmetric) $4N \times 4N$
matrix, representing the impurity potential (superconducting order
parameter).  The form (\ref{eq:def V}) of the potential ${\cal V}$
ensures that the Hamiltonian ${\cal H}$ obeys particle-hole 
symmetry:
${\cal H}=-\gamma_1{\cal H}^{{\rm T}}\gamma_1$.\cite{AltlandZirnbauer}
In addition, ${\cal H}$ (and hence ${\cal V})$ may obey time-reversal
invariance ${\cal H} = {\cal T}{\cal H}^* {\cal T}^{-1}$, with 
${\cal T}=i\tau_1\otimes\sigma_2$.

In Ref.\ \onlinecite{BFGM} it was assumed that the potentials 
$\Delta$ and $v$ in Eq.\ (\ref{eq:def V}) are Gaussian white-noise 
potentials with the 
same variance. It was for this choice of disorder that the
``critical'' conductance statistics was found. 
Motrunich {\em et al.} have argued that this choice is special: 
As soon as the variances of $\Delta$ and $v$ are no longer equal, 
the conductance will decay exponentially with length $L$. 
In addition, they showed that if a term proportional
to $v_F \tau_2/2 l_s$ is added to the Hamiltonian, which corresponds to a 
staggering of the hopping amplitude in a lattice version of the
problem,\cite{BMSA} the localization length
is finite even if $v$ and $\Delta$ have the same variance. 

Below we shall analyze the effects of a different variance of $\Delta$
and $v$ and of staggering on the localization length, and discuss
how these effects change with the thickness of the wire. We will 
first address the diffusive regime, where an
analytical treatment is possible, and then the localized regime,
where we support our arguments with numerical simulations.

{\em Different variances of $\Delta$ and $v$.} 
For simplicity, we assume that time-reversal symmetry 
and spin-rotational symmetry are completely broken. Hence we take
the potentials $v$ and $\Delta$ as Gaussian white-noise, 
with vanishing means and with variances
\begin{eqnarray}
  \langle v_{ij}^{\vphantom{*}}(x) v_{kl}^*(y) \rangle &=&
  \frac{v_F^2}{8 N l_{v      }} 
  \delta_{ik}^{\vphantom{*}} \delta_{jl}^{\vphantom{*}}\, 
  \delta(x-y),
  \label{eq:avg} \\
  \langle \Delta_{ij}^{\vphantom{1}}(x) \Delta_{kl}^*(y)
   \rangle &=&
  \frac{v_F^2}{8 N l_{\Delta}} 
  \left(
  \delta_{ik}^{\vphantom{*}} \delta_{jl}^{\vphantom{*}} 
  -
  \delta_{il}^{\vphantom{*}} \delta_{jk}^{\vphantom{*}}
  \right) \delta(x-y).
  \nonumber 
\end{eqnarray}
Here $l_v$ and $l_{\Delta}$ are the mean free paths
for scattering from potential disorder and
from (fluctuations of) the superconducting pair potential $\Delta$,
respectively. 
When $l_{\Delta} = l_{v}$, the exact solution
of Ref.\ \onlinecite{BFGM} applies. We will refer to 
this special case as the pure class D and consider the 
generic case $l_{\Delta} \neq l_{v}$ as a point in the
crossover between the standard unitary ensemble and
class D. (The standard unitary ensemble corresponds to the absence
of superconducting correlations, $l_{\Delta} = \infty$.)
The combined mean free path $l$ for scattering from 
either $v$ or $\Delta$ is $l^{-1} = l_{v}^{-1} + 
l_{\Delta}^{-1}$. In the special case $l_{\Delta} = l_{v}$,
i.e., in the pure class D, the ensemble average dimensionless
conductance $\langle g \rangle$ in the diffusive regime $L \ll N l$
reads\cite{BFGM}
\begin{equation}
\label{eq:WL}
  \langle g \rangle = 
  \frac{4Nl}{L} 
  + \frac{1}{3} 
  + {\cal O}\left(\frac{L}{N l}\right).
\end{equation}
For $L \gg N l$, the average conductance decays algebraically,
whereas the ``typical conductance'' $\exp(\langle \ln g \rangle)$ 
decreases proportional to $\exp[-(2L/\pi N l)^{1/2}]$,\cite{BFGM}
\begin{eqnarray}
  && \langle g \rangle  =  \sqrt{\frac{8 N l}{\pi L}}, \qquad
\label{eq:localizedD}
\langle \ln g \rangle = -\sqrt{\frac{2 L}{\pi N l}}.
\end{eqnarray}
For the standard unitary ensemble $l_{\Delta} = \infty$
one has $\langle g \rangle = 4 N l/L + {\cal O}(L/N l)$ in the
diffusive regime, while
in the localized regime $\langle g \rangle \propto e^{-L/8 N l}$, 
$\langle \ln g \rangle = -L/2 N l + {\cal O}(1)$.\cite{BeenakkerReview}

Let us now consider the generic case $l_{v} \neq l_{\Delta} < \infty$,
i.e., a point in the crossover between the standard unitary
ensemble and the pure class D. At a fixed but large 
$l_{\Delta} \gg l_{v}$, this same crossover can also take place as 
a function of the wire length $L$. In that case, short $L$ corresponds
to the standard unitary ensemble, while class D statistics might be
found for large $L$. To find
the relevant length scale for that crossover we calculate the 
average conductance in the diffusive regime. Following the method
of Ref.\ \onlinecite{MBF} we derive scaling equations for the averages 
of certain traces of transmission and reflection matrices $t$ and $r$
of the wire. For $l_{\Delta}, L \gg l$
and up to corrections of relative order $(L/N l)$
the relevant scaling equations read
\begin{eqnarray}
  && 
  {4 N l}\, \partial_{L} 
  \left\langle \mbox{tr}\, t^{\dagger} t \right\rangle =
  - \left\langle \mbox{tr}\, t^{\dagger} t \right\rangle^2 
  \nonumber \\ 
  &&
  \hphantom{
  {4 N l}\, \partial_{L} 
  \left\langle \mbox{tr}\, t^{\dagger} t \right\rangle=} 
  + 2 \left[ 
  \left\langle 
  \mbox{tr}\, 
  \left(
  r^{\dagger} r t^{\dagger} t 
  -  \gamma_3 r^{\dagger} \gamma_3 r t^{\dagger} t 
  \right)
  \right\rangle 
  \right],
%%%%%%%%%%%%%%%%%%%%%%%%%%%%%%%%%%%%%%%%%% end first equation 
  \nonumber \\  && \nonumber\\
  && 
  {4 N l}\, \partial_{L} 
  \left\langle 
  \mbox{tr}\, \left(1 - \gamma_3 r^{\dagger} \gamma_3 r\right) 
  \right\rangle =
  - \left\langle 
    \mbox{tr}\, (1 - \gamma_3 r^{\dagger} \gamma_3 r)
    \right\rangle^2 
  \nonumber \\ 
  &&
  \hphantom{
  {4 N l}\, \partial_{L} 
  \left\langle 
  \mbox{tr}\, \left(1 - \gamma_3 r^{\dagger} \gamma_3 r\right) 
  \right\rangle=}
  + (\xi_{\Delta}/l)^2,
%%%%%%%%%%%%%%%%%%%%%%%%%%%%%%%%%%%%%%%%%% end second equation 
  \label{eq:set} \\  && \nonumber\\
  && 
  {4 N l}\, \partial_{L} 
  \left\langle 
  \mbox{tr}\, \gamma_3 r^{\dagger} \gamma_3 r t^{\dagger} t 
  \right\rangle = 
  -2 \left\langle 
  \mbox{tr}\, \left(1 - \gamma_3 r^{\dagger} \gamma_3 r\right) 
  \right\rangle 
  \nonumber \\ 
  &&
  \hphantom{
  {4 N l}\, \partial_{L} 
  \left\langle 
  \mbox{tr}\, \gamma_3 r^{\dagger} \gamma_3 r t^{\dagger} t 
  \right\rangle=} 
  \times
  \left\langle 
  \mbox{tr}\, \gamma_3 r^{\dagger} \gamma_3 r t^{\dagger} t
  \right\rangle, 
%%%%%%%%%%%%%%%%%%%%%%%%%%%%%%%%%%%%%%%%%% end third equation 
  \nonumber \\  && \nonumber\\
  && 
  {4 N l}\, \partial_{L} 
  \left\langle 
  \mbox{tr}\, r^{\dagger} r t^{\dagger} t
  \right\rangle = 
  -2 \left\langle \mbox{tr}\, t^{\dagger} t \right\rangle 
  \left\langle \mbox{tr}\, r^{\dagger} r t^{\dagger} t  \right\rangle,
%%%%%%%%%%%%%%%%%%%%%%%%%%%%%%%%%%%%%%%%%% end fourth equation 
  \nonumber %\\ && 
\end{eqnarray}
where we abbreviated
$\xi_{\Delta} = \sqrt{l l_{\Delta}/8}$ and $\gamma_3$ is the
Pauli matrix in particle-hole grading. Solving Eqs.\ (\ref{eq:set}) we
find the average conductance $\langle g \rangle = \langle \mbox{tr}\, 
t^{\dagger} t \rangle$ for $l_{\Delta},L \gg l$,\cite{ballistic}
\begin{eqnarray}
  \langle g \rangle &=& \frac{4 N l}{L} + \frac{1}{3} +
  \frac{\xi_{\Delta}^2}{L^2}
  - \frac{\xi_{\Delta}}{L} \coth\frac{L}{\xi_{\Delta}}
  + {\cal O}\left( \frac{L}{N l}\right).
  \label{eq:WL2}
\end{eqnarray}
The length scale $\xi_{\Delta}$
marks the crossover between
the standard unitary class and class D: For $L \ll \xi_{\Delta}$ we 
recover the result $\langle g \rangle = {4 N l/ L} + 
{\cal O}(L/N l)$
of the standard unitary ensemble, while for $L \gg \xi_{\Delta}$ Eq.\
(\ref{eq:WL2}) simplifies to the class D result (\ref{eq:WL}). As
Eq.\ (\ref{eq:WL2}) is only valid in the diffusive regime $L \ll N l$, 
one can only expect to observe the class D result (\ref{eq:WL}) if
$\xi_{\Delta}$ is smaller than $N l$. In the thick-wire limit this
poses no real constraint, and we conclude that, in the thick-wire
limit, any finite mean free
path $l_{\Delta}$ will eventually give rise to a weak localization
correction to the conductance according to class D if the wire is
sufficiently long.

To find what happens in the localized regime we now compare the
crossover length scale $\xi_{\Delta}$ derived above
and the localization length $4N l$ in the
standard unitary class. If $\xi_{\Delta} \gg N l$, the
wave functions are still governed by the unitary
class on the scale $N l$, not by class D, when localization
sets in. Hence there is no room for class D 
physics to play a role and the localization length is that
of the standard unitary class. On the other hand, 
if $\xi_{\Delta} \ll N l$, the statistics of wave functions and
conductance is given by that of class D at scales $\xi_{\Delta}$ and
beyond. Therefore if $\xi_{\Delta} \ll N l$ we expect that the 
wave function and conductance statistics for $L \gg N l$ is critical, 
as in class D. 
If $N l$ and $\xi_{\Delta}$ are comparable, the conductance
statistics in the localized regime is best characterized as
in the ``crossover'' between the two universality classes:
The localization length is expected to 
be larger than $4N l$, but still finite. 

In order to make these statements more precise, we look at the 
normalized inverse localization length,
\begin{equation}
  \chi(\sigma) = - \frac{\langle \ln g \rangle}{\sigma},
  \qquad \sigma=\frac{L}{2 N l}.
\label{eq: definition chi}
\end{equation}
In the standard unitary ensemble, one has $\chi \to 1$ as 
$L\to \infty$ with $N$, $l_{\Delta}$, and $l_{v}$ held fixed; 
critical conductance statistics implies $\chi(\infty) = 0$. A finite
value $0 < \chi(\infty) < 1$ corresponds to exponentially localized
wave functions, but with a localization length that is larger
than in the standard unitary ensemble. Taking the limit 
$L \to \infty$ at fixed values of $N$, $l_{\Delta}$, and $l_{v}$
corresponds to a point somewhere in the crossover
between the standard unitary ensemble 
and class D if $l_{\Delta} \neq l_{v}$.
As such a choice of parameters 
corresponds to a {\em finite} distance from the pure 
class D, we expect $\chi(\infty) > 0$ in this case, corresponding 
to exponentially localized
wave functions. This agrees with the claim of Motrunich 
{\em et al.},\cite{Motrunich} who found critical statistics
at the point $l_{\Delta} = l_{v}$ only. 

In the thick-wire limit, however, the picture is different. 
The thick-wire limit is defined as the simultaneous limit $L \to 
\infty$ and $N \to \infty$, keeping $\sigma = L/2N l$ fixed. 
The limit $N \to \infty$ arises as a semiclassical limit,
when the Fermi wavelength is much smaller than the diameter of the
wire, or by increasing wire diameter. In the latter case, the diameter
should not exceed the transverse localization length.
Based on our calculation for the diffusive regime, we surmise that in
the thick-wire limit
the crossover between the standard unitary class and class D 
is described by a two-parameter
scaling function $\chi^*(\sigma,\alpha)$,
which is obtained from $\chi$ by taking the limit $L, N \to 
\infty$, at fixed values of the scaling variables $\sigma$ and 
$\alpha = 2 N l/\xi_{\Delta}$.\cite{foot3}
The second scaling variable $\alpha$ is the ratio of
the localization length $4N l$ in the standard unitary class 
and twice the diffusive crossover length scale $\xi_{\Delta}$
(the proportionality constant is arbitrary as long as it is of order 1).
As a function of $\alpha$,
$\chi^*$ interpolates between the standard unitary
ensemble for $\alpha = 0$
and the critical conductance statistics of class 
D if $\alpha \to \infty$.
Note that taking
the thick-wire limit at any finite and fixed ratio of $\xi_{\Delta}$
and $l$ (or, alternatively, of $l_{\Delta}$ and $l$)
corresponds to the case $\alpha = \infty$, 
i.e., automatically realizes class D. Thus, in the sense of the scaling
limit defined above, the standard unitary fixed point is ``unstable,''
whereas the class D fixed point is ``attractive.'' 
It is only by allowing $\xi_{\Delta}
/l$ to scale at least as fast as $N$ that one can probe non-class-D
physics in the thick-wire limit. 

\begin{figure}

\includegraphics[width=\columnwidth]{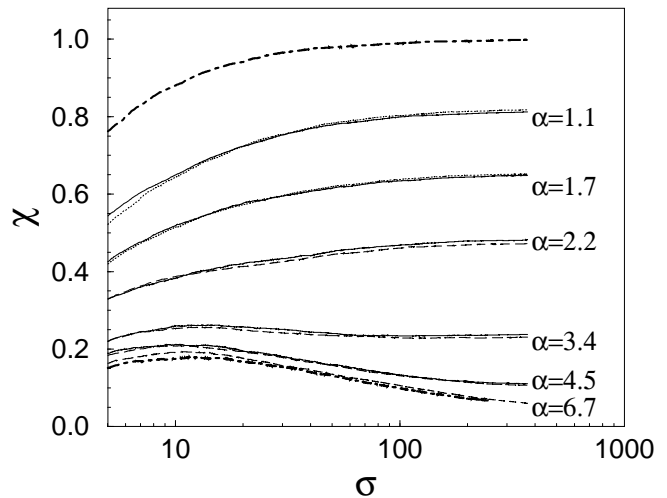}

\caption{\label{fig:1}
Normalized inverse localization length 
$\chi = -\langle \ln g \rangle/\sigma$ versus $\sigma = L/2Nl$. 
Solid curves are for $l_{\Delta} = 201\, l$ 
($\xi_{\Delta} = 5.01\, l$) and $N=4,6$, from top to bottom. 
Dotted and dashed curves are for
$l_{\Delta}=101\, l$ ($\xi_{\Delta} = 3.6\, l$) with $N=2,3,4,6,8$,
and $l_{\Delta} = 26\, l$ ($\xi_{\Delta} = 1.8\, l$) with
$N=2,3,4,6$, respectively. Thick dash-dot curves at the top 
and bottom are for the standard unitary class ($l_{\Delta}^{-1}=0$) and 
$N=1$ and for the pure class D ($l_{\Delta} = 2 l$) and 
$N=3$, respectively. Curves with equal values of the scaling
variable $\alpha = 2N l/\xi_{\Delta}$ but different 
$\xi_{\Delta}/l$ or $N$ almost coincide; the remaining small 
differences between them are attributed to finite-$N$ effects.
Corresponding values of $\alpha$ are listed to the right of 
each set of curves with the same $\alpha$.
}
\end{figure}

We verified the above scenario with numerical simulations of $\chi$
as a function of $\sigma$ for a
discretized version of the model (\ref{eq:cal H}). The discretized
model is obtained from Eq.\ (\ref{eq:cal H})
by the replacement of the continuous potential 
${\cal V}(x)$ by the sum $\sum_{j} {\cal
V}_{j} \delta(x - a j)$, where $a$ is a microscopic length scale.
The delta functions $\delta(x-y)$ in Eq.\
(\ref{eq:avg}) are replaced by Kronecker deltas $a^{-1} \delta_{jj'}$.
In Fig.\ \ref{fig:1}, results are shown for different 
values of $l_{\Delta}/l$ and for different $N$. We took an
average over $> 2000$ realizations of the disorder potential.
Our conclusions are confirmed by the numerical data:
First, curves for the same value of $\alpha = 2 N l/\xi_{\Delta} =
4 N(2 l/l_{\Delta})^{1/2}$ approach a well-defined limit for
large $N$, confirming that $\alpha = 2 N l/\xi_{\Delta}$ is indeed
the true crossover parameter. 
Second, in the crossover between the two classes,
one still observes localized behavior, but with a significantly
enhanced localization length, which increases as $\alpha$ is
increased.

{\em Staggering.} Staggering is described by the addition of a term
$v_F \tau_2/2l_s$ to the Hamiltonian (\ref{eq:cal H}). Without the
potential term ${\cal V}$ in Eq.\ (\ref{eq:cal H}), the effect of
staggering would be to create a gap of size $v_F/l_s$ around the
Fermi energy $\varepsilon = 0$. The effect of a random 
potential ${\cal V}$ is to create states inside this gap. 
Below we address the localization properties
of such states at the Fermi level $\varepsilon=0$ if the random
potential ${\cal V}$ is of the form (\ref{eq:def V}) with nonzero
superconducting correlations $\Delta$ and without spin-rotational
invariance and time-reveral symmetry. 
Based on a 
combination of analytical and numerical arguments, we argue that
eigenstates at the Fermi level remain critical (as in the absence
of staggering), provided that
the staggering is sufficiently weak [see Eq.\ (\ref{eq:stagcrit})
below] and the thick-wire limit is taken.

To study the effect of staggering on the $\varepsilon=0$ eigenstates
of the Hamiltonian (\ref{eq:cal H}), we first 
investigate the conductance in the diffusive regime. Following
the method of Ref.\ \onlinecite{MBF}, we construct scaling equations
for the moments
$R_{n,m} = \langle \mbox{tr}\, r^{n} r^{\dagger m} \rangle$. 
To leading order in $N$, they read
\begin{widetext}
\begin{eqnarray}
  \partial_L R_{n,m} &=&
  \frac{n}{l_s} (R_{n+1,m} - R_{n-1,m})
  +
  \frac{m}{l_s} (R_{n,m+1} - R_{n,m-1})
  -
  \frac{2 (n+m)}{l}
  R_{n,m} 
  \nonumber \\ &&+
  \frac{1}{4 N l} 
  \sum_{p=0}^{n-1} \sum_{q=0}^{m-1} 
  \left[R_{p,q} (R_{n-1-p,m-1-q} + R_{n-p,m-q})
  + R_{n-p,m-q} (R_{p+1,q+1} + R_{p,q}) \right]
  \nonumber \\ &&- 
  \frac{1}{Nl} 
  \sum_{p=1}^{n-1}
  (n-p) R_{p,0} R_{n-p,m}
  - \frac{1}{Nl} 
  \sum_{q=1}^{m-1}
  (m-q) R_{0,q} R_{n,m-q}. \label{eq:fnm}
\end{eqnarray}
\end{widetext}
Although the scaling equations
(\ref{eq:fnm}) do not form a closed set, a solution can be constructed
as a series expansion in $l/l_{s}$. For sufficiently 
weak staggering,
\begin{equation}
  l_s > l,
  \label{eq:stagcrit}
\end{equation}
and for lengths $L \gg l$, the perturbation series in $l/l_{s}$
truncates after the second order, and one finds the simple solution
\begin{equation}
  \langle g \rangle =
  \langle \mbox{tr}\, (1 - R_{1,1})\rangle = \frac{2}{\sigma}
 + {\cal O}(1),
\end{equation}
where $\sigma$ is now defined as
\begin{equation}
  \sigma = \frac{L}{2N l [1 - (l/l_s)^2]}.
  \label{eq:leff}
\end{equation}
Hence, in the thick-wire limit, staggering causes a renormalization 
of the scaling length $\sigma$, but does not remove the existence of a 
diffusive regime. If  staggering exceeds the
critical strength (\ref{eq:stagcrit}), there is no diffusive regime
in which $\langle g \rangle$ is inversely proportional to wire length
at $\varepsilon = 0$. The existence of a critical
staggering strength is consistent with
numerical simulations of periodic-on-average single-channel quantum 
wires by Deych {\em et al.}\cite{Deych}

\begin{figure}

\includegraphics[width=\columnwidth]{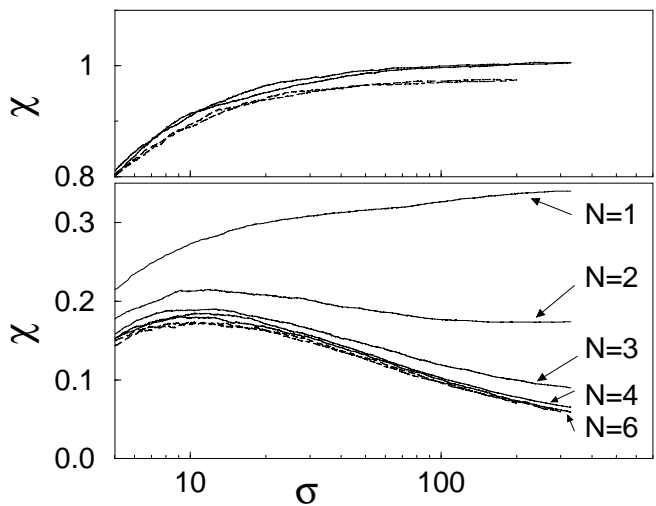}

\caption{\label{fig:2}
Normalized inverse localization length $\chi
= - \langle \ln g \rangle/\sigma$ versus scaled wire length 
$\sigma$ with and 
without staggering. With staggering, $\sigma$ is given by
Eq.\ (\protect\ref{eq:leff}), without staggering $\sigma = L/2 N l$.
Upper panel shows results for normal quantum wires
($l_{\Delta} = \infty$). Solid curves are for
$l_v = l = 0.80 l_s$ and for $N=4$, $6$; dashed
curves are without staggering and $N=4$, $6$. The slight
difference (less than $5\%$) between curves with and without 
staggering is attributed to a small deviation from Eq.\ 
(\protect\ref{eq:leff}) due to the finite disorder strength
in the simulations. As all curves for different $N$ collapse, 
no single curves are labeled. Lower panel shows
results for superconducting quantum wires with $l_{\Delta}
= l_{v}$. Solid curves are for $l_v = l_{\Delta} = 2l = 1.60 l_s$ 
and $N=1,2,3,4$, and $6$. 
Dashed curves are without staggering and $N=3$, $N=4$. They
almost coincide with the curve with staggering and
$N=6$ and are not labeled individually. 
}
\end{figure}

Based on this result for the diffusive regime, we propose that
in the simultaneous limit $N \to \infty$, $L \to \infty$ the 
normalized inverse localization length 
$\chi= -\langle \ln g \rangle/\sigma$ 
approaches the same scaling limit 
$\chi^*(\sigma)$ as in the absence of staggering, if $\sigma$ is
given by Eq.\ (\ref{eq:leff}).
In Fig.\ \ref{fig:2} we show the result of 
numerical simulations for the discretized version of Eq.\ (\ref{eq:cal
H}) with staggering. We have compared the cases $l_v/l_{\Delta} =
1$ or $0$, with or without (weak) staggering and find that for sufficiently
large $N$ all curves $\chi$ versus $\sigma$ with the same symmetry of
the Hamiltonian collapse. That is, curves with $l_v/l_{\Delta} = 1$ 
(or, in fact, any finite ratio $l_v/l_{\Delta}$) approach
the class D result, while curves with $l_v/l_{\Delta} = 0$ 
approach the result of the standard unitary class. Small $N$ curves
are slightly different, which again indicates the absence of universality
when a diffusive regime is absent.

In conclusion, our analytical and numerical results confirm the 
claim of Motrunich {\em et al.}\ that the conductance statistics
in the localized regime is not determined by symmetry alone for
a superconducting quantum wire with a small number $N$ of propagating
channels at the Fermi level.\cite{Motrunich}
Such universality is only achieved for ``thick'' quantum wires with 
$N \gg 1$. Which situation is relevant
depends on the physical system at hand. For quasiparticle states at
the edges of superconducting planes in high-$T_c$ superconductors, 
the nonuniversal scenario of small $N$ may be more 
applicable, while we expect that the universal scenario of large
$N$ is more appropriate for dirty unconventional superconducting wires 
or normal-metal wires in the proximity of a superconductor.
Possible relevant perturbations to the critical behavior in
the thick-wire limit of the superconducting class D are
disorder sufficiently strong to induce transverse localization,
restoration of the spin-rotation symmetry, or
breaking of the particle-hole symmetry (e.g., by tuning the chemical 
potential away from the the Fermi level or by electron-electron
interactions). All of these perturbations are expected to restore 
the localized phase at large length scales.

We would like to point out that the universality --- or lack thereof ---
for the localization properties in superconducting quantum wires
is not different from the case of normal-metal wires. In both cases,
universality is expected in the thick-wire limit only. Or, in the
language of scaling flow: In
both cases, the fixed point with less symmetry is attractive, but 
the flow towards the fixed point is only complete if the thick-wire 
limit is taken. In that 
respect, note that for the crossover between the
standard unitary class and the superconducting class D considered in
this paper, the superconducting class is the class of less symmetry. 
That fact constitutes a qualitative
difference between class D and the three chiral classes, which also
may exhibit critical conductance statistics for wire length $L \gg 
N l$.\cite{BMSA} In contrast to the superconducting class D, 
the chiral classes have very 
high symmetry, which makes them rather unstable and which will render
observation of the critical conductance statistics in these classes
more difficult.

We would like to thank K.\ Damle and I.\ A.\ Gruzberg for discussions. 
This work was supported in part by the NSF under Grant No.\ DMR 0086509, 
by the Sloan and Packard foundations (P.W.B.), and 
by a Grant-in-Aid for Scientific Research on Priority Areas (A) from the
Ministry of Education, Culture, Sports, Science and Technology  
(No. 12046238) (A.F.).
P.W.B.\ and A.F.\ thank the Institute for Theoretical Physics
in Santa Barbara for its hospitality during the final stages of
this work. C.M.\ thanks the Yukawa Institute for Theoretical Physics
in Kyoto for its hospitality during the final stages of this work.


\begin{thebibliography}{ourpaper}
\bibitem{ballistic}
For $l_{\Delta} \approx l_{v}$ or $l_{\Delta} \ll l_{v}$ the 
class D result (\ref{eq:WL}) is obtained already for wire lengths
$L$ of the order of the mean free path.
\bibitem{foot3}
Note that the limit $\sigma \to \infty$
of the scaling function $\chi^*(\sigma,\alpha)$ is a different
quantity than the limit $N \to \infty$ of the
Lyapunov exponent, which is defined by taking the
limit $L \to \infty$ \textit{before} the limit $N \to \infty$
in Eq.\ (\ref{eq: definition chi}).
\end{thebibliography}
\end{document}